\documentclass[twocolumn]{aastex631}

\usepackage{savesym}
\savesymbol{tablenum}

\usepackage{graphicx}%
\usepackage{multirow}%
\usepackage{amsmath,amssymb,amsfonts}%
\usepackage{amsthm}%
\usepackage{mathrsfs}%
\usepackage[title]{appendix}%
\usepackage{xcolor}%

\usepackage[version=4]{mhchem}
\usepackage{hyperref}
\usepackage[separate-uncertainty=true]{siunitx}
\usepackage{float}
\usepackage{lipsum}

\renewcommand{\d}[1]{\ensuremath{\operatorname{d}\!{#1}}}
\DeclareSIUnit\bar{bar}
\DeclareSIUnit\AU{AU}
\DeclareSIUnit\dex{dex}
\DeclareSIUnit\erg{erg}
\DeclareSIUnit\day{day}
\DeclareSIUnit\year{yr}

\newcommand{\GPCC}{\gram\per\centi\meter\cubed}

\newcommand{\amu}{\gram\per\mole}
\newcommand{\IW}{\text{IW}}
\newcommand{\ppmw}{\text{ppmw}}
\newcommand{\Hppmw}{\text{H}_\ppmw}

\usepackage{newfloat}
\DeclareFloatingEnvironment[name={Supplementary Figure}]{suppfigure}
\usepackage{sidecap}
\sidecaptionvpos{figure}{c}

\begin{document}

\title{Volatile-rich evolution of molten super-Earth L\,98-59\,d}

\correspondingauthor{Harrison Nicholls}
\email{harrison.nicholls@ast.cam.ac.uk}

\author[0000-0002-8368-4641]{Harrison Nicholls}
\affiliation{Department of Physics, University of Oxford, Oxford, United Kingdom}
\affiliation{Institute of Astronomy, University of Cambridge, Cambridge, United Kingdom}

\author[0000-0002-3286-7683]{Tim Lichtenberg}
\affiliation{Kapteyn Astronomical Institute, University of Groningen, Groningen, The Netherlands}

\author[0009-0008-8739-0932]{Richard D. Chatterjee}
\affiliation{Department of Physics, University of Oxford, Oxford, United Kingdom}
\affiliation{School of Physics and Astronomy, University of Leeds, Leeds, United Kingdom}
\

\author[0000-0003-1521-5461]{Claire Marie Guimond}
\affiliation{Department of Physics, University of Oxford, Oxford, United Kingdom}
\affiliation{Department of Earth and Planetary Sciences, ETH Z\"urich, Z\"urich, Switzerland}

\author[0009-0009-5036-3049]{Emma Postolec}
\affiliation{Kapteyn Astronomical Institute, University of Groningen, Groningen, The Netherlands}

\author[0000-0002-5887-1197]{Raymond T. Pierrehumbert}
\affiliation{Department of Physics, University of Oxford, Oxford, United Kingdom}

\begin{abstract}
Small low-density exoplanets are sculpted by strong stellar irradiation, but their primordial compositions and subsequent evolution are still unknown. Two often-considered scenarios hold that they formed with rocky interiors and \ce{H2}-He atmospheres (`gas-dwarfs'), or alternatively with bulk compositions dominated by \ce{H2O} phases (`water-worlds'). Here, we constrain the possible range of evolutionary histories linking the birth conditions of low-density super-Earth L\,98-59 d to recent observations using a coupled atmosphere-interior evolutionary model. We find that the observations can be explained by in-situ photochemical production of \ce{SO2} in an \ce{H2} background, indicative of a chemically-reducing mantle and substantial ($>1.8\text{ mass\%}$) early sulfur and hydrogen content, inconsistent with both the gas-dwarf and water-world scenarios. L\,98-59 d’s interior comprises a permanent magma ocean, allowing long-term retention of volatiles within its mantle over billions of years, consistent with California-Kepler Survey trends. Our analysis reveals an evolutionary pathway in which planets host volatile-rich atmospheres sustained by long-term magma ocean degassing, shaped by secular cooling, atmospheric erosion and photochemistry. Internal and environmental processes contribute to the observed diversity of super-Earth and sub-Neptune exoplanets.
\end{abstract}



\section{Introduction}
\label{sec:body}

Exoplanets with radii between $\sim1.5$ and $\sim4.0$\,R$_\oplus$ are abundant and amenable to characterisation using current instruments; yet they have no Solar-System analogues \citep{lichtenberg_review_2025}. This small-planet regime includes the super-Earth and sub-Neptune populations, between which there exists an apparent paucity in the surveyed population known as the `radius valley', although it remains unclear how these exoplanet populations originate \citep{fulton_california-_2018, david_cks_2021}. Two separate scenarios are often considered to explain the properties of these planets \citep{valencia_diversity_2025, Zeng2019}: (i) \textit{gas-dwarfs} where both populations jointly form with large \ce{H2}-dominated envelopes and then diverge into two distinct populations due to long-term atmospheric escape processes \citep{owen_review_2019, lopez_born_2017}, and (ii) \textit{water-worlds} where super-Earths and sub-Neptunes differ in \ce{H2O} content obtained from formation relative to the ice-line \citep{mousis_water_2020, luque_density_2022,Lacedelli2022,Burn2024}. Combinations of both scenarios may explain the emergence of sub-Neptunes and super-Earths on a population level, but not on an individual-planet basis \citep{bean_nature_2021, rogers_road_2025}. An apparent over-abundance of young sub-Neptune planets may represent a reservoir of future super-Earths, which if true, intimately links these two proposed populations across the radius valley \citep{fernandes_signature_2025, david_cks_2021}.

L\,98-59 hosts three transiting exoplanets \citep{demangeon_warm_2021, kostov_l9859_2019}. Hubble Space Telescope observations of the innermost sub-Venus planet rule out a low mean-molecular-weight (MMW) primary atmosphere \cite{zhou_hubble_2022}, but recent JWST transit spectroscopy has indicated an \ce{SO2}-rich atmosphere \citep{belloarufe_evidence_2025}, raising the possibility of an S-rich planetary system. Hubble observations of the outermost transiting planet (L\,98-59\,d) have also ruled out a pure \ce{H2} atmosphere \citep{zhou_hubble_2023}. Yet L\,98-59\,d's exceptionally low bulk-density, recently estimated as \SI{2.2}{\GPCC} (with mass 1.64\,M$_\oplus$ and radius 1.627\,R$_\oplus$), is inconsistent with a pure rock-and-iron composition; instead consistent with it hosting substantial atmosphere-forming volatiles  \citep{cadieux_detailed_2025, demangeon_warm_2021}.  Retrievals on its transit spectrum point to an atmosphere of sulfur-bearing volatiles (\ce{H2S}, \ce{SO2}) within a background of \ce{H2}. Free-chemistry models fit at $5.6\sigma$ with a MMW of $9.18_{-2.41}^{+2.51}$ \SI{}{\amu} and an \ce{H2S} volume mixing ratio of $10^{-0.74_{-0.49}^{+0.14}}$ \citep{gressier_hints_2024}. However, thermochemical equilibrium models fit with a MMW of $32.13_{-8.31}^{+1.50}$ \SI{}{\amu} at a lower significance of only $2.7\sigma$. Retrievals also place weak constraints on \ce{H2O}, \ce{CO2}, and \ce{CH4} \citep{banerjee_atmospheric_2024}. Furthermore, the presence of \ce{H2S} in the atmosphere of L\,98-59\,d has been reaffirmed through high resolution spectrography -- using ground-based IGRINS observations -- that provides evidence for \ce{H2S} with a Bayes factor of 390 \citep{cheverall_ground_2026}. Sitting proximate to the radius valley, L\,98-59\,d stands out among the low-mass exoplanets as prime for tracing the physics of rocky planet formation and evolution \citep{parc_from_2024}.

Here, we constrain the plausible set of evolutionary pathways for L\,98-59\,d by numerically linking its birth conditions to recent estimates of its bulk density $\rho_p$ and atmospheric composition. For this comparison we primarily adopt $\rho_p=\SI{3.45}{\GPCC}$ previously estimated by \citet{rajpaul_doppler_2024} and the corresponding upper-atmospheric composition constraints from JWST \citep{gressier_hints_2024,banerjee_atmospheric_2024}. An alternative recent estimate of a smaller $\rho_p=\SI{2.2}{\GPCC}$ is simultaneously explored (\cite{cadieux_detailed_2025}, Table~1). We use the \textsc{proteus} modelling framework to create a large grid of self-consistent coupled atmosphere-interior evolutionary simulations to compare with observational constraints \citep{lichtenberg_vertically_2021, nicholls_redox_2024, nicholls_convective_2025, nicholls_tidal_2025}. \textsc{proteus} calculates the evolving thermochemical state of the planet's interior and atmosphere over time. From a molten-start, we simultaneously model bottom-up mantle crystallisation, tidal dissipation, outgassing and escape of CHNOS volatiles, atmospheric blanketing and energy transport, and stellar evolution. Our radiative-convective atmosphere implements a real-gas EOS, correlated-$k$ radiative transfer, and allows the formation of deep radiative layers \citep{innes_runaway_2023, nicholls_convective_2025}. Stellar XUV fluxes are used to calculate the rate of volatile loss (see Methods) using an energy-limited hydrodynamic escape parametrisation \citep{lehmer_rocky_2017}. Our main grid of 900 models has four axes: initial volatile endowment (proxied as hydrogen relative to the total mantle mass, $1000 \le \Hppmw \le 16000$), bulk S/H mass ratio, magma ocean oxygen fugacity, and total planet mass. The grid domain spans a range of post-accretion scenarios (Supplementary~Figure~\ref{fig:si_outgas}) as planets are thought to vary substantially in their initial amount of volatiles  \citep{Zeng2019,Venturini2020} in addition to variations their mantle redox state \citep{lichtenberg_review_2025, righter_Redoxvar_2016} which controls the solubility of volatiles in silicate melt and chemical speciation throughout the planet \citep{gaillard_redox_2022, krissansen_erosion_2024}. Models terminate at the planet's estimated age of \SI{4.94(1.44)}{\giga\year} or at mantle solidification; we do not model secondary volcanic outgassing from magmatism during solid-state mantle convection. \textsc{proteus} simulation end-points are  compared with observational constraints on the planet's density \citep{rajpaul_doppler_2024} and atmospheric MMW \citep{gressier_hints_2024} to infer L\,98-59\,d's thermal and compositional history.

\section{Methods}
\label{sec:met}

\subsection{Modelling framework} 
\label{sec:met_framework}

We use \textsc{proteus} to simulate the time evolution of L\,98-59 up to the present day \citep{lichtenberg_vertically_2021, nicholls_redox_2024, nicholls_tidal_2025}. \textsc{proteus} is a self-consistently coupled modular simulation framework which links the \textsc{spider} interior evolution model \citep{bower_numerical_2018, bower_retention_2022} and the \textsc{agni} radiative-convective atmosphere model \citep{nicholls_convective_2025, nicholls_agni_2025}. The atmospheric composition is determined by  equilibrium thermochemistry and the solubility of volatiles into the magma ocean \citep{gaillard_redox_2022, sossi_solubility_2023}. Volatiles are assumed to partition between the partially-molten mantle and overlying atmosphere at solubility equilibrium, subject to mass conservation, and the two-phase dynamics and crystallisation of the mantle at each point in time. The total planetary inventory of CHNS volatiles is initialised via our model parameters; we assume that the stellar nebula has already dispersed, and no primary-accreted atmosphere remnant is considered. However, the planet's volatile inventory continually evolves due to removal by hydrodynamic escape, and redox reactions of O with Fe in the mantle, set by the temperature-dependent iron-w\"ustite buffer reaction. Partitioning of volatiles into solid phases \citep{hier_origin_2017, sim_volatile_2024, guimond_mantle_2023} and volcanic outgassing \citep{liggins_growth_2022} are not included in the model, as all cases of interest retain magma oceans; volcanic mass fluxes associated with a solidified interior are several orders of magnitude below those from efficient magma ocean degassing \citep{lichtenberg_review_2025,gaillard_diverse_2021,guimond_low_2021}. The mantle material itself is chemically inert. 
 
The pure-\ce{MgSiO3} molten mantle is initialised on an adiabat, and evolved over time by \textsc{spider} according to the energy fluxes at each radial level. The initial specific entropy is set to \SI{3150}{\joule\per\kelvin\per\kilo\gram}; the planet's subsequent evolution is insensitive to this exact value. Mixing-length theory convection, solid-liquid phase separation, grain gravitational settling, radiogenic heating, core cooling, and tidal heating are modelled self-consistently. Tidal heating profiles are calculated using \textsc{lovepy} \citep{hay_io_2020} under a Maxwell viscoelastic rheology; the depth-dependent density, shear modulus, bulk modulus and shear viscosity are calculated from the melt fraction profiles solved by \textsc{spider} using empirical scaling relations \citep{kervazo_solid_2021}. Application of more realistic rheological parametrisations would yield increased tidal heat fluxes, and thus hotter and more molten interiors \citep{renaud_rheology_2017, farhat_tides_2024}. As in several previous studies \citep[e.g.][]{bower_retention_2022, lichtenberg_vertically_2021, hamano_lifetime_2015, schaefer_magma_2016, nicholls_redox_2024,krissansen_erosion_2024} we do not spatially resolve energy transport \textit{within} the metallic core, and instead fix its bulk density and heat capacity to representative Earth-like \citep{DZIEWONSKI1981297} values of \SI{10.738}{\GPCC} and \SI{880}{\joule\per\kelvin\per\kilo\gram} following \citep{bower_numerical_2018}.  A pure Fe core of the same relative radius would be denser in L\,98-59\,d compared to Earth, due to compression by mantle overburden pressure. Although the compression would be offset by incorporation of lighter elements (SNCHO, etc.) into the core, decreasing its density compared to pure Fe \citep{Luo2024, Hirose2021}. Sensitivity tests assess the impact of this interior structure assumption on our results (Supplementary~Figure~\ref{fig:si_sens}). Heat is transported across the core-mantle-boundary following \citep{bower_linking_2019}. \textsc{spider} uses the \citet{wolf_equation_2018} equation of state for \ce{MgSiO3} melt; mantle density is independent of dissolved volatile content, which can be neglected in the modelled regime \citep{DornLichtenberg2021}.  The core radius is nominally fixed at 55\% of the interior radius (see sensitivity test in Supplementary~Figure~\ref{fig:si_sens}), consistent with the relative size of Earth's core \citep{lodders_planetary_1998}, although alternative estimates on the planet's density permit theoretically constraining the planet's core size (Supplementary~Figure~\ref{fig:si_density_updated}).  The radius $R_\text{int}$ of the planet's interior (mantle plus core) is calculated using \textsc{spider} by solving for its hydrostatic structure for a given interior mass $M_\text{int} = M_p - M_\text{atm}$. For a planet mass $M_p=2.14M_\oplus$, we obtain $M_\text{core} = \SI{3.9e24}{\kilo\gram} = 0.66 M_\oplus$ and $M_\text{mant} = \SI{8.85e24}{\kilo\gram} = 1.48 M_\oplus$. The interior radius $R_\text{int}$ is held constant throughout each simulation; thermal contraction and mantle solidification does not chiefly affect the volume of silicate mantles \citep{boley_fizzy_2023, unterborn_pressure_2019, lichtenberg_review_2025}. 
 
Energy transport through an optically-thick atmosphere regulates the planet's energy balance. At each time-step, our radiative-convective atmosphere model \textsc{agni} calculates the net atmospheric energy flux for the outgassed chemical composition, instellation flux, and top-of-mantle temperature. The radiation fluxes at each atmosphere level are calculated using \textsc{socrates} \citep{edwards_studies_1996, amundsen_radiation_2014} correlated-$k$ two-stream ($\theta_\text{zen}=54.74^\circ$) radiative transfer with 48 spectral bands, including Rayleigh scattering and non-grey basaltic surface emissivity. Schwarzschild-stable mixing-length theory \citep{joyce_mlt_2023} is applied to calculate the convective energy flux. An energy-conserving atmospheric solution is obtained using the Newton-Raphson method with line-search. The incorporation of a real gas EOS formulation into \textsc{agni} is presented in Supplementary~Figure~\ref{fig:si_struct}.
 
Evolution of stellar radius, bolometric luminosity, and photometric emission is modelled using \textsc{mors} \citep{johnstone_active_2021}. We take L\,98-59 to have a mass of $0.273 M_\odot$ and present-day rotation rate of 80.9\,days in our main grid of models \citep{engle_living_2023}. \textsc{mors} uses a two-shell rotational evolution calculation with empirically-derived scalings to estimate the bolometric and XUV emission from the star over its lifetime. Using this established code, the estimated X-ray luminosity at the present day is consistent with XMM Newton observations of L\,98-59 \citep{behr_muscles_2023,nicholls_tidal_2025}. We use the L\,98-59 spectrum from the MUSCLES Extension \citep{behr_muscles_2023}. Stellar quantities are updated throughout the simulations; changes in stellar flux are thus reflected in the atmospheric temperature profile, albedo, and escape rate. 
 
All simulations are initialised at a planet age of $t_\mathrm{ini}=$\SI{50}{\mega\year} relative to the formation of the system under the assumption that boil-off has completed \citep{tang_cpml_2024} and further delivery processes are negligible compared to the physics and timescales simulated \citep{lichtenberg_geophysical_2022}. Simulations terminate when the simulated age reaches the \SI{4.94}{\giga\year} estimated age of L\,98-59 \citep{engle_living_2023} or when the mantle solidifies (mass melt-fraction $\Phi < 0.5\%$). Cases consistent with the observations are those with a planet bulk-density $\rho_p$ comparable ($\pm1\sigma$) with an estimated density of \SI{3.45}{\GPCC} in our main results \citep{rajpaul_doppler_2024}. A recent radial velocity and transit-timing reanalysis has yielded a even-lower bulk density of \SI{2.2}{\GPCC} \citep{cadieux_detailed_2025}; our comparison with the previous higher-estimate of $\rho_p$ (Figure~\ref{fig:density_blame}) thereby yields conservatively lower estimates on total volatile content. We compare our models to these new estimates in Supplementary~Figure~\ref{fig:si_density_updated}). $\rho_p = 3M_\text{p}/4 \pi R_p^3$ depends on the total planet mass $M_p$ and the observable radius $R_p$ that would effectively be probed by transit measurements. $M_p$ includes the mantle, metallic core, dissolved volatiles, and atmosphere. The modelled observable radius $R_p$ is determined by the radius corresponding to the 20\,mbar pressure level \citep{lehmer_rocky_2017}, to which our results are not sensitive.
 
\subsection{Energy-limited escape}
\label{sec:met_escape}
Atmospheric escape is modelled within the \textsc{proteus} framework using the classic energy-limited formulation of photoevaporative mass-loss,
\begin{equation}
    \dot{M}_\text{esc}(t) = \frac{\eta \pi [R_\text{XUV}(t)]^3 F_\text{XUV}(t)}{G M_p},
    \label{eq:EL_escape}
\end{equation}
which assumes that stellar XUV flux $F_\text{XUV}$ absorbed by an optically thick disk of radius $R_\text{XUV}$ lifts mass [kg/s] at a rate $\dot{M}_\text{esc}$ out of the gravitational potential-well with efficiency $\eta$, at each point in time $t$. The time-dependent $F_\text{XUV}(t)$ is calculated with \textsc{mors} \citep{johnstone_active_2021}. $R_\text{XUV}(t)$ is determined from the atmospheric temperature structures calculated by \textsc{agni} (see Supplementary Information), taking $P_\text{XUV} = \SI{5}{\pascal}$  \citep{lehmer_rocky_2017}.
 
An atmosphere composed of hydrogen and high-MMW elements will experience compositional fractionation at varying degrees depending on the XUV irradiation, collisional coupling between species, and gravity. In the limit of rapid hydrodynamic escape, and where the high-MMW elements compose a substantial fraction of the atmosphere by mass, hydrodynamic simulations from \citep{Johnstone_2020} have shown that the thermosphere is expected to escape in bulk without fractionating. Here, we explore high-MMW atmospheres exposed to XUV fluxes greater than a hundred times present Earth levels, so we adopt this approximation of no thermospheric fractionation. Including fractionation in our model of escape would not change our major conclusions, as it would only enhance the S/H ratio.

We adopt a 10\% escape efficiency \citep{lehmer_rocky_2017, owen_review_2019}, which agrees with simulations from \citep{yoshida_escape_2024} that account for molecular line cooling (Supplementary~Figure~\ref{fig:si_maxesc}). The elemental composition of the escaping gas is set equal to that of the outgassed atmosphere, which would be photodissociated and partially photoionised on outflow. To calculate the loss rate of each element from the planet's \textit{total} volatile inventory, we proportion the bulk escape rate by the \textit{atmospheric} mass mixing ratios of each element. So, while the escape process itself is treated as non-fractionating, it acts upon an \textit{atmospheric} elemental composition different to the \textit{bulk} planetary composition. Volatile loss thereby fractionates the planet's elemental inventories as a result of the interior-atmosphere partitioning of HCNS elements.

\section{Results and discussion}

\subsection{A volatile-rich birth}
\label{sec:res_evolve}

\begin{figure*}[htbp]
    \centering
    \includegraphics[width=0.9\textwidth, keepaspectratio]{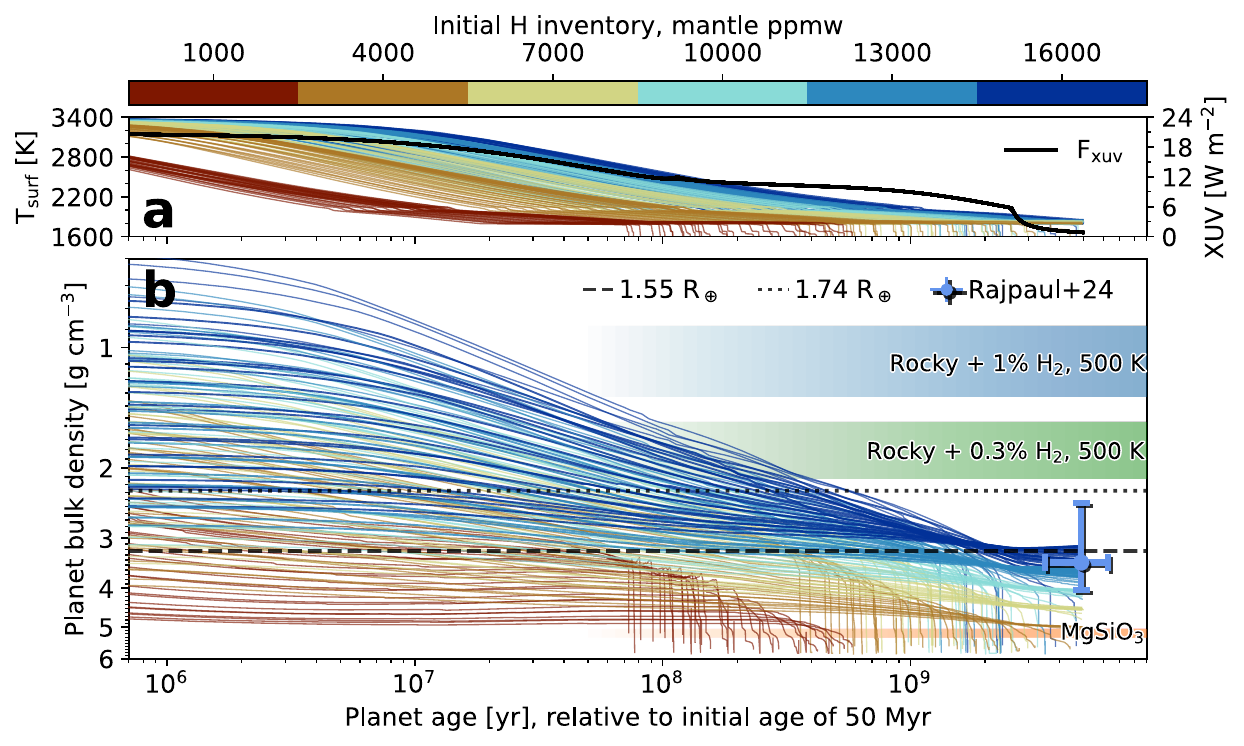}%
    \caption{\textbf{Modelled planetary bulk-density over time} (panel a). The initial hydrogen inventory of each bulk-density evolution track is shown by the line colour. Blue/green shaded regions are reference densities for a planet of this mass \citep{Zeng2019}. Black dashed/dotted lines demarcate edges of the radius valley (at this planet mass) for low mass stars \citep{ho_shallower_2024}. The region $\pm1\sigma$ compatible with the estimated bulk-density of this planet is indicated by the blue errorbar \citep{rajpaul_doppler_2024, demangeon_warm_2021}. Panel b: evolution of surface temperature (coloured) and stellar XUV energy flux (black). For visual clarity, this figure only shows cases with S/H=8.}
    \label{fig:density}
\end{figure*}

Modelled evolutionary tracks (Figure~\ref{fig:density}b) show that the bulk-density $\rho_p$ of L\,98-59\,d has changed substantially over its lifetime. For larger initial volatile inventories (blue lines), $\rho_p$ is typically less than \SI{2}{\GPCC} during the first several Myr of evolution. The planet's observable effective radius ($R_p$; see Methods) initially exceeds the $1.7 R_\oplus$ value often taken as the transition between the super-Earth and sub-Neptune regimes. Some models remain deep within the sub-Neptune regime ($R_p \ge 1.7 R_\oplus$) for \SI{950}{\mega\year}. Radiative energy losses drive initial cooling and atmospheric contraction, causing the planet's radius to shrink below the upper edge of the radius valley (dotted line, Figure~\ref{fig:density}b). Irreversible loss of CNHOS volatiles through XUV-driven photoevaporation dominates the planet's bulk density evolution at later times up to the present. Scenarios which fall within estimates of the present-day $\rho_p$ \citep{rajpaul_doppler_2024} are all initially volatile-rich: $\Hppmw\gtrsim13000$ with CHNOS volatiles comprising $\gtrsim 1.8 \%$ of the planet's mass.
 
The mantle solidifies under volatile-poorer formation scenarios (Figure~\ref{fig:density}b, maroon lines) from \SI{70}{\mega\year} onwards. Surface cooling and near-complete escape of degassed volatiles leads to an increase in $\rho_p$ towards that of a rocky composition within \SI{650}{\mega\year}. These volatile-stripped scenarios are not compatible with the observed $\rho_p$ despite initial H-contents likely larger than those of the inner Solar System planets \citep{krijt_chemical_2023}. These cases, applicable to high-density exoplanets not modelled here, reflect suggestions that airless exoplanets may be the stripped interiors of evolved sub-Neptunes \citep{lopez_born_2017, owen_evap_2017}.

\begin{figure}[htbp]
    \centering
    \includegraphics[width=\linewidth, keepaspectratio]{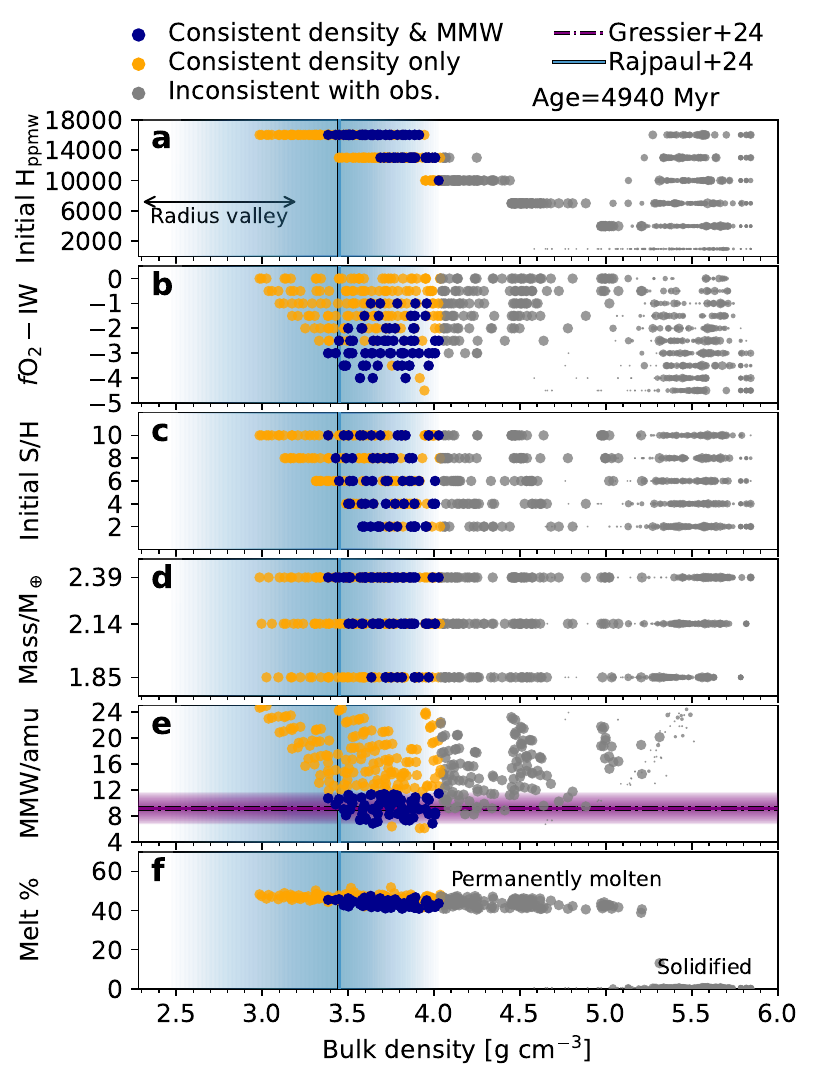}%
    \caption{\textbf{Projection of planetary bulk-density $\rho_p$ against several variables}. Scatter points represent end points of simulations from our grid, projecting the calculated bulk density against other variables: H inventory from formation (panel a), mantle oxygen fugacity (panel b), S/H ratio from formation (panel c), total planet mass (panel d), atmospheric MMW (panel e), and mantle melt fraction (panel f). Point sizes represent the age of the planet at simulation end-points; largest scatter point size corresponds to the present day. Vertical blue line and shaded region ($\pm1\sigma$) highlights the observationally estimated $\rho_p$. Horizontal purple line show estimates on MMW (with $\pm1\sigma$ shaded) and \ce{H2S} from free chemistry retrievals. Blue points are consistent with the observed density and MMW, orange points are only consistent with the observed density.}
    \label{fig:density_blame}
\end{figure}

Which scenarios are compatible with the observations? Figure~\ref{fig:density_blame} plots modelled $\rho_p$ at ages corresponding to the present-day. Values of $\rho_p$ are projected over the grid axes (panels a, b, c, d) and other variables. All cases compatible with the $\rho_p$ estimated by \citet{rajpaul_doppler_2024} (orange points) exhibit MMWs from 7~to~\SI{24}{\amu} (panel e); a spread wider than the $\pm1\sigma$ range of $9.18^{+2.51}_{-2.41} \SI{}{\amu}$ (purple line) derived with JWST \citep{gressier_hints_2024}. Simultaneously applying the density and MMW constraints yields the smaller set of blue points.

Formation with $\Hppmw < 10000$ can be readily ruled-out on the basis of $\rho_p$ alone (grey points in Figure~\ref{fig:density_blame}). Here, we compare our model against the maximum estimate of $\rho_p$ in the literature, so these conclusions are insensitive to the quoted uncertainty in $\rho_p$. Larger escape efficiencies, or lower bulk densities such as recently estimated by \citet{cadieux_detailed_2025}, necessitate even volatile-richer birth scenarios to satisfy the present-day $\rho_p$ (Supplementary~Figure~\ref{fig:si_density_updated}). 

Figure~\ref{fig:density_blame}b shows that magma ocean oxygen fugacity $f\ce{O2}$ (expressed in log-units relative to the iron-w\"ustite buffer) must be between $\IW-4$ and $\IW-1$ in order to reproduce the estimated $\rho_p$ and MMW. More oxidising conditions favour gas speciation towards O-bearing species, increasing the atmospheric MMW. A modest MMW ($\sim\SI{9}{\amu}$) lowers the atmosphere's propensity for escape through a decreased radius. Larger MMW atmospheres are either over-dense or inconsistent with the JWST transit depths (or both). The full range of \textit{initial} S/H mass ratio modelled is compatible with the observed $\rho_p$. Variations in the planet's dry mass itself has only minor impact on modelled $\rho_p$ compared to the other variables which drive changes in its radius $R_p$ (Supplementary~Figure~\ref{fig:si_sens}). 

All outcomes compatible with the estimated small bulk density have a substantial atmosphere that induces a strong greenhouse effect. Alongside tidal-heating of its interior, these models show that L\,98-59\,d presently retains a permanent magma ocean (melt fraction $\Phi\sim 45\%$, Figure~\ref{fig:density_blame}f). A solidified mantle is ruled-out by $\rho_p$, without invoking the MMW constraint, making this inference insensitive to uncertainties on the JWST observations and retrievals. The narrow range of mantle melt fractions corresponding to the observed $\rho_p$ is a physical outcome; mantle viscosity increases strongly at $\Phi\lesssim45\%$ \citep{costa_model_2009}, which makes energy transport through the planet's interior inefficient, and thereby allows atmospheric blanketing and tidal heating to keep a permanent magma ocean \citep{nicholls_tidal_2025}. In comparison, modelled thin atmospheres lead to mantle solidification (Figure~\ref{fig:density_blame}).

\subsection{Thermal contraction and photoevaporation}
\label{sec:res_diagnose}

\begin{figure}[htbp]
    \centering
    \includegraphics[width=\linewidth, keepaspectratio]{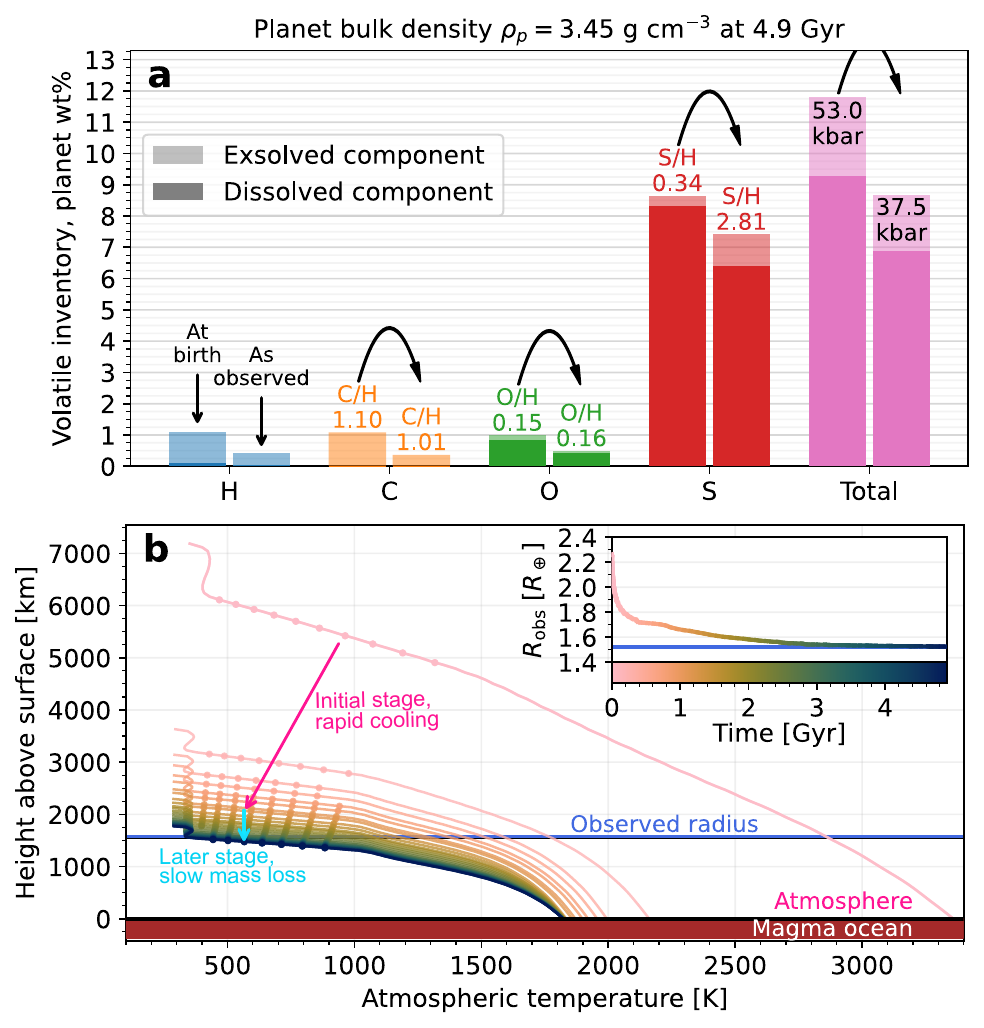}%
    \caption{\textbf{Volatile loss and atmosphere contraction over time, through two stages of evolution}. Bar heights in panel (a) highlight the total loss of volatiles between planet birth and observation, as percentages relative to total planet mass. Lighter and darker bar opacities indicate partitioning between the atmosphere and interior, respectively. Atmosphere elemental mass ratios relative to H are annotated. Panel (b) visualises the evolving atmospheric temperature profile, with an initial stage of rapid contraction due to cooling, followed by a later-stage of slower contraction due to mass loss. Dotted markers indicate convective regions. Profile line colours correspond to time, relative to model initialisation (colourbar). The colour bar is mapped to the x-axis of the inset, which plots radius $R_p$ as a function of time. }
    \label{fig:casestudy}
\end{figure}

Many modelled scenarios \textit{initially} have large radii on the sub-Neptune side of the radius valley (Figure~\ref{fig:density}b). To understand how volatile loss and thermal contraction drive the marked decrease in $R_p$ over time, passing through the radius valley, we study a case compatible with the observations in detail. The case study in Figure~\ref{fig:casestudy} shows L\,98-59\,d losing 26\% of its total volatile inventory whilst retaining a large surface pressure of \SI{30}{\kilo\bar} (pink bars) at the present-day. Some volatiles are degassed during this time due to partial mantle solidification; the majority of the planet's volatiles remain dissolved in the magma ocean. The atmosphere S/H mass ratio increases by a factor of 8.2 because sulfur, which is more soluble than H in silicate melts, is degassed as the magma ocean crystallises \citep{namur_mercury_2016, gaillard_redox_2022}. Photoevaporation also drives a relative enhancement in bulk S-content due to the favourable interior-partitioning of sulfur.

Thermal contraction explains large initial changes in $\rho_p$. The surface cools from 3360 to 1830\,K over the first 1.4\,Gyr, corresponding to effective radius deflating from $>2.2R_\oplus$ to $\sim1.74R_\oplus$ during that period, despite volatile degassing simultaneously driving a net \textit{increase} to the surface pressure (Supplementary~Figure~\ref{fig:si_driver}). The evolving atmospheric temperature-height profile is visualised in Figure~\ref{fig:casestudy}b. While early atmospheric contraction is an established theoretical behaviour of \ce{H2}-dominated envelopes, important in regulating the rate of photoevaporative volatile loss \citep{kubyshkina_coupling_2020, lopez_understanding_2014}, here we show that interior cooling is the primary driver of early radius contraction for envelopes for which MMW increases substantially over time. The atmosphere has a deep radiative layer, up to $\sim\SI{5000}{\kilo\meter}$ thick, above which it undergoes dry convection and is then weakly inverted. High temperatures make radiative diffusion efficient at transporting energy \citep{pierrehumbert_book_2010}. Radiative layers in the deep atmosphere decrease the lapse rate compared to an adiabat, acting to inflate the atmospheric radius \textit{for a given surface temperature} during the planet's evolution. Prior modelling largely assumed adiabatic or isothermal atmospheres, neglecting this important impact of energy transport on atmospheric structure. 

After initial thermal contraction, tidal forces and atmospheric blanketing together sustain a deep magma ocean on L\,98-59\,d \citep{nicholls_tidal_2025} while escape has reduced its radius down to its observed value. Importantly, the magma ocean buffers the atmosphere through volatile dissolution into the melt and equilibrium thermochemistry near the surface.  It has been previously suggested for \ce{H2O} \citep{DornLichtenberg2021} and N-compounds \citep{shorttle_k218b_2024} that retention of volatiles in deep magma oceans buffers atmospheric escape over deep time, while analogues of Mercurian lavas have established that reducing melts may take up large S inventories \citep{namur_mercury_2016}. Dissolution of sulfur (Figure~\ref{fig:casestudy}a) enables the planet's S inventory to be retained across gigayear timescales and enhances atmospheric S/H over time, as H atoms dominate the escaping outflow. This scenario aligns with Venus/Earth/Mars models in which their present atmosphere masses were largely supplied by magma ocean degassing \citep{elkins_MagmaOce_2012, hirschmann_Magmaoce_2012} with later volcanic contributions in uncertain amounts; Archean proxies suggest that early outgassing established most of Earth's present atmospheric mass  \citep{marty_Nitrogen_2013} although its chemistry has changed substantially \citep{Halliday2023}.

\subsection{Photochemical production of \texorpdfstring{SO$_2$}{SO2}}
\label{sec:res_photochem}

\begin{figure}[htbp]
    \centering
    \includegraphics[width=\linewidth, keepaspectratio]{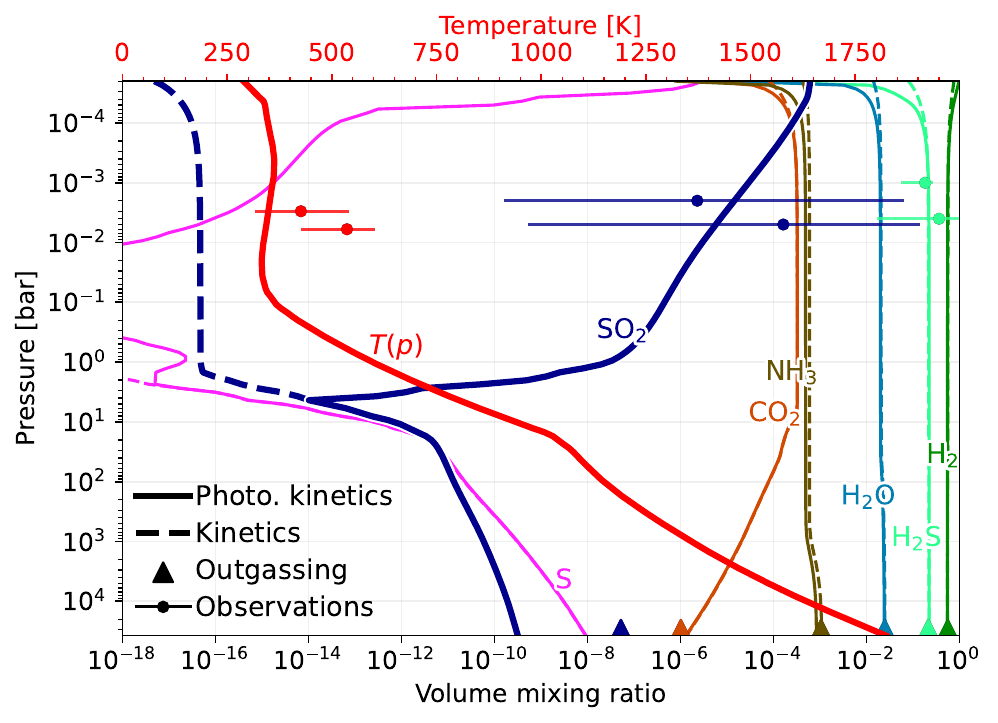}%
    \caption{\textbf{Atmospheric composition and temperature profiles}. Solid lines plot the volume mixing ratios for a selection of gases calculated with \textsc{vulcan}'s SNCHO photochemical kinetic network. Dashed lines plot mixing ratios calculated without photoreactions. Scatter points show the median-estimates for photospheric chemical abundances (blue, lime) and temperatures (red) retrieved by \citet{banerjee_atmospheric_2024} and \citet{gressier_hints_2024}. Scatter point error bars represent $\pm1\sigma$ ranges on the JWST retrieval posteriors. We assume a modest eddy diffusion coefficient $K_{zz}$ of \SI{1e5}{\centi\meter\squared\per\second} and use the radiative-convective temperature solution obtained by \textsc{agni} (thick red line).  }
    \label{fig:photochem}
\end{figure}

Formation of \ce{SO2} from \ce{H2S}, in the presence of \ce{OH} radicals from photolysis \ce{H2O}, can explain the detections of \ce{SO2} in L\,98-59\,d's atmosphere. This mechanism produces \ce{SO2} within the hot-Jupiter WASP-39b \cite{tsai_photochemically_2023, powell_sulfur_2024}, and \ce{SO2} -- in the presence of percentage-levels of \ce{H2O} -- has been detected in a Neptune-mass exoplanet \citep{gressier_hatp26b_2025}. We infer the formation of \ce{SO2} by making a comparison between \textsc{vulcan} photochemical kinetics models and free-chemistry retrievals \citep{gressier_hints_2024}. Figure~\ref{fig:photochem} plots atmospheric mixing ratios for our case study under three chemical paradigms: \textsc{vulcan} SNCHO photochemical kinetics (solid lines), kinetics without photochemistry (dashed lines), and the isochemical volatile outgassing as applied during our evolutionary calculations (triangle markers). The mixing ratio of \ce{SO2} (solid blue line) increases with altitude for $p<\SI{6}{\bar}$, while the equivalent case without photochemistry (dashed blue line) quenches at negligible abundance. Photochemical production of \ce{SO2} is necessary to raise its abundance to that consistent with the JWST observations (error-bars). Lime-coloured profiles show that \ce{H2S} abundance is consistent with the observational constraints under all three paradigms; \ce{H2S} is thermochemically favoured to carry sulfur in the \ce{H2}-rich background. 

Detections of \ce{SO2} cannot be explained by surface outgassing \citep{nicholls_tidal_2025}, which would require \ce{SO2} to be transported into the observable upper atmosphere without being thermochemically reduced. Transport of \ce{SO2} (\SI{64}{\amu}) would have to take place in the absence of deep convection (Figure~\ref{fig:casestudy}b) and therefore only by diffusion through the \ce{H2} background. This physically unlikely scenario instead points to the formation of \ce{SO2} in-situ. Production of \ce{SO2} requires the presence of \ce{H2O} to form \ce{OH} radicals; a strongly water-free composition would be physically incompatible with the detection of \ce{SO2}. Our photochemical models predict \ce{H2O} abundances compatible with the wide observational posterior $\log_{10}\chi_{\ce{H2O}} = -6.63^{+4.22}_{-3.51}$ derived by \citet{banerjee_atmospheric_2024}. However, an alternative analysis of the same JWST spectrum \citep{gressier_hints_2024} yields an upper-limit of $\log_{10}\chi_{\ce{H2O}}<-5$. The abundance of \ce{H2O} is poorly constrained by both analyses since \ce{H2O} absorption overlaps with multiple features in a common wavelength region. Future observations should robustly check for percent-level \ce{H2O} in the atmosphere of L\,98-59\,d to test for the presence of these photochemical processes. Our comparison with JWST data in Figure~\ref{fig:photochem} shows that photochemistry actively shapes the observables of super-Earths, dependent on the chemical interactions linking the deep interiors of planets to their upper-atmospheres. 

\section{Conclusions}
\label{sec:conclude}

Our modelled evolution pathways connect the observed bulk-density of L\,98-59\,d to a narrow range of conditions at birth. Conditions compatible with observations are those in which the planet initially had $\gtrsim100$ times the estimated hydrogen content of the early Earth mantle \citep{krijt_chemical_2023,wang_elements_2018}. The majority of L\,98-59\,d's H and C are stored in its atmosphere, whilst its S remains primarily dissolved within a reducing magma ocean. Hydrogen may also incorporate into the planet's metallic core, allowing additional H storage in-bulk and lower core densities \citep{wade_core_2005} H concentrations up to $\sim7300$\,ppmw at high pressures may allow the majority of Earth's H to be stored in its core \citep{peslier_water_2017}. So, although we infer a small core for L\,98-59\,d, our estimated inventories represent lower-limits on volatiles delivered during formation.  Fractionation of planetary HCNS inventories between interiors and atmospheres presents an opportunity for testing whether small, low-density exoplanets have underlying magma oceans through measurements of their atmospheric composition \citep{shorttle_k218b_2024, DornLichtenberg2021, kite_exchange_2016}.  

L\,98-59\,d's radius is smaller than that of a prototypical sub-Neptune, although had it been observed at earlier age $\lesssim \SI{1.4}{\giga\year}$ we may have confidently labelled it as one (Figure~\ref{fig:density}b). Its transition from the sub-Neptune to super-Earth regime occurs only after several Gyr of thermo-compositional evolution, suggesting that even sub-Neptunes with ages $\gtrsim \SI{1}{\giga\year}$ may exist in a transient state towards becoming what would later be classified as super-Earths. This Gyr-scale contraction has been suggested to explain radius-age trends in the Kepler survey \citep{david_cks_2021} as an alternative to the `loss and revival' mechanism \citep{kite_exoplanet_2020}. 

L\,98-59\,d currently has a radius commensurate with a super-Earth planet in the vicinity of the radius valley, but its low bulk-density necessitates that it comprises substantial volatiles. A modestly large atmospheric MMW means that it is neither an evolved gas-dwarf with a primitive atmosphere sourced from protoplanetary disk gas; nor is a water-world scenario required to explain its current bulk-properties. Moreover, forming a substantially water-dominated atmosphere would require oxidising conditions incompatible with the MMW inferred from JWST transmission spectroscopy, pointing to geochemical conditions in the interior of L\,98-59\,d that are unknown from the Solar System. Instead, global-scale disequilibrium evolution driven by thermal contraction, magma ocean degassing, tidal heating, photochemistry, and photoevaporative mass-loss have together continuously shaped the radius and atmospheric composition of L\,98-59\,d over its lifetime.

\pagebreak

\section{Supplementary information}

\subsection{Observed planetary parameters}
\label{sec:si_params}

Table~1 outlines the most recent estimates for the orbital and bulk parameters of L\,98-59\,d, derived from the NASA Exoplanet Archive. It has a zero-albedo equilibrium temperature of \SI{416}{\kelvin} \citep{demangeon_warm_2021}. Note that the planet radius $R_p$ represents the observable radius of the planet at the level probed photometrically during transit (primary eclipse); $R_p$ is not equal to the radius of the surface.

The orbital eccentricities estimated for L\,98-59\,b/c/d have previously merited an investigation into the stability of this system. Taking median estimates for $e$ and $a$ on each planet yields an unstable orbital configuration; stability is readily achieved by adjusting eccentricity of only planet c from 0.147 to 0.103, while leaving the other planets unperturbed \citep{demangeon_warm_2021}.


{
\setlength{\extrarowheight}{.5em}
\begin{table*}
\begin{tabular}{l|lllll} 
Parameter & \citeauthor{cadieux_detailed_2025} & \citeauthor{demangeon_warm_2021} & \citeauthor{rajpaul_doppler_2024} & \citeauthor{luque_density_2022} & \citeauthor{cloutier_character_2019} \\ \hline 

$a$ [\SI{0.01}{\AU}]     & $4.94\pm0.16 $            & $4.86^{+0.18}_{-0.19}$    & --                        & --                     & $5.06\pm0.02$   \\

Eccentricity             & $0.006^{+0.007}_{-0.004}$ & $0.074^{+0.057}_{-0.046}$ & $0.098^{+0.027}_{-0.096}$ & --                     & $<0.09$                \\ 

$R_p/R_\oplus$           & $1.627\pm0.041$           & $1.521^{+0.119}_{-0.098}$ & $1.521^{+0.119}_{-0.098}$ & $1.58\pm0.08$          & $1.57\pm0.14$          \\ 

$M_p/M_\oplus$           & $1.64\pm0.07$             & $1.94 \pm 0.28$           & $2.14^{+0.25}_{-0.29}$    & $2.31^{+0.46}_{-0.45}$ & $2.31^{+0.46}_{-0.45}$ \\ 

$\rho_p$ [\SI{}{\GPCC}]  & $2.2\pm0.2$               & $2.95^{+0.79}_{-0.51}$    & $3.45^{+0.25}_{-0.29}$    & $3.17^{+0.85}_{-0.73}$ & $3.3^{+1.3}_{-0.9}$   \\

Star period [day]& $77.5\pm1.6 $             & $80.9^{+5.0}_{-5.3}$      & $58\pm20$                 & --                     & $78\pm13 $   \\

Star mass [M$_\odot$]    & $0.2923\pm0.0067 $        & $0.273\pm0.030 $          & --                        & --                     & $0.312\pm0.031$   \\

\end{tabular}%
\caption{\textbf{Literature estimates for pertinent parameters of L\,98-59\,d and its star}. We tabulate semi-major axis $a$, orbital eccentricity, planet radius $R_p$, planet mass $R_p$, bulk-density $\rho_p$, and stellar rotation period. These quantities were obtained through a combined re-analysis of HARPS and ESPRESSO datasets\citep{demangeon_warm_2021,cloutier_character_2019,rajpaul_doppler_2024}, alongside TESS and JWST photometric transit-timing variations \citep{cadieux_detailed_2025}. The extreme $\pm1\sigma$ range on the bulk-density across these five works is 2.0~to~\SI{4.6}{\GPCC}.}
\label{tab:si_l9859d}
\end{table*}
}

\subsection{Modelling atmospheric structure}
\label{sec:si_struct}

Atmospheric thermal $T(P)$ and structural $r(P)$ evolution is important in setting the atmospheric escape rates of young sub-Neptune planets after formation \citep{lopez_understanding_2014, kubyshkina_coupling_2020}. The small MMW of \ce{H2} and high temperatures mean that sub-Neptune atmospheres are initially inflated, allowing for efficient `boil-off' mass-loss \citep{tang_cpml_2024}. This propensity for early volatile loss could extend to lower mass planets \citep{lehmer_rocky_2017}, especially those orbiting pre-main sequence stars with large XUV luminosities during their saturated phase \citep{johnstone_active_2021}.

An accurate representation of atmospheric structure is important for comparing models to observations, and for accurately modelling atmospheric escape rates. Several previous studies on early atmospheric structural evolution and escape have assumed convective temperature profiles \citep[e.g.][]{rogers_road_2025,cherubim_oxidation_2025, krissansen_erosion_2024, lehmer_rocky_2017, wordsworth_water_2013, rogers_most_2015}. However, it has been shown that convectively-stable deep radiative layers can form \citep{innes_runaway_2023,selsis_cool_2023, nicholls_convective_2025}. Compared to an adiabatic profile, the shallow lapse rates ($\d{T}/\d{p}$) of dry radiative layers yield cooler surfaces and hotter upper-atmospheres at radiative equilibrium. For a \textit{given} surface temperature $T_s$, radiative layers yield larger atmospheric temperatures. 
 
Here, the atmosphere is defined on a pressure grid of 40 layers between the top $P_t$ and the surface $P_s$. The surface pressure evolves over time, calculated at each step by our volatile outgassing scheme \citep{nicholls_redox_2024, shorttle_k218b_2024}. $P_t$ is fixed at \SI{1}{\pascal}. Assuming that the atmosphere is hydrostatically self-supported, we relate the pressure and radius
\begin{equation}
    dP = -\rho(P,T) g(r) dr,
    \label{eq:hydrostatic}
\end{equation}
where $\rho(P,T)$ is the mass-density of the gas at pressure $P$ and temperature $T$, $g(r)$ is the gravitational acceleration, and $r$ is the radial distance from the planet centre. Equation \ref{eq:hydrostatic} is integrated from $P_s$ to $P_t$. The radius $r=R_\text{int}$ and gravity at the surface are calculated by \textsc{spider} as described above. The first improvement made for this work is to express the gravity $g(r)$ at each layer as
\begin{equation}
    g(r) = \frac{G}{r^2} M(r)
\end{equation}
where $M(r)$ is the \textit{total} mass enclosed within $r$, accounting for the gravitational attraction of the upper atmosphere layers to the atmosphere layer below (as well as to the planet interior). Self-attraction is important for these atmospheres which comprise percentage-fractions of the total mass. \textsc{agni} now also implements an RK4 scheme to integrate Equation \ref{eq:hydrostatic}, using 30 sublevels within each of the atmosphere layers.
 
The second improvement here is the implementation of a real-gas EOS, used to evaluate $\rho(P,T)$ in Equation \ref{eq:hydrostatic} and adiabatic lapse rate $\nabla_\text{ad}$. Whenever the atmospheric composition or temperature is updated, the densities $\rho_j$ of each volatile component $j$ are re-evaluated separately using their appropriate EOS, and then combined using Amagat's additive volume law to obtain the total mass-density $\rho$ at each layer:
\begin{equation}
    \frac{1}{\rho(P,T)} = \sum_j \frac{1}{y_j \rho_j(P,T)},
    \label{eq:amagat}
\end{equation}
where $y_j$ are mass mixing ratios. This construction permits non-ideal behaviour for each gas component, but assumes that species mix without interacting. Ideal mixing has long been applied in stellar physics \citep{dorman_eos_1991, saumon_eos_1995, baraffe_new_2015} and for \ce{H2}-dominated planetary atmospheres \citep{kubyshkina_coupling_2020,paxton_mesa_2011}. Lab experiments and DFT simulations have shown that ideal mixing is robust and provides the flexibility necessary for modelling mixtures at extreme conditions \citep{magyar_eos_2014, magyar_ethane_2015, bradley_experimental_2018}. The EOS for water is interpolated from \citep{haldemann_aqua_2020} and the EOS for hydrogen is interpolated from \citep{chabrier_eos_2019}. The Van der Waals EOS is applied for {CO2}, \ce{CH4}, \ce{CO}, \ce{N2}, \ce{NH3}, \ce{SO2}, \ce{H2S}, and \ce{O2}. Any other gases are treated as ideal.

To validate our updated atmospheric structure calculation, here we apply \textsc{agni} to a range of binary \ce{H2}-\ce{H2O} mixtures. The atmosphere of L\,98-59\,d is thought to be mainly composed of \ce{H2}, with volatiles of higher MMW mixed-in \citep{gressier_hints_2024,banerjee_atmospheric_2024,demangeon_warm_2021}. We use \ce{H2O} as a proxy for heavy species here because of its particular potential for non-ideal behaviour. Supplementary~Figure~\ref{fig:si_struct} plots modelled atmospheric structures for a range of \ce{H2}-\ce{H2O} mixtures (line colour), EOS formulations (rows), and temperature profile assumptions (line style). Solid lines are radiative-convective models where an energy-conserving solution is obtained for a given instellation, and given surface temperature $T_s$ (3000 K) and pressure $P_s$ (1000 bar). Dashed lines are defined by the dry adiabat until they intersect the \ce{H2O} saturation curve: an assumption typically applied in the literature. Note that these profiles are calculated for a \textit{fixed} $T_s$, which represents the early phase of atmospheric evolution above a hot magma ocean, and not at radiative equilibrium (which occurs after Gyr-scale time evolution).
 
Panels (a) and (b) of the figure show that the radiative-convective profiles (solid lines) are not always on the corresponding adiabat (dashed lines). Stellar radiation heats the upper layers, while thermal radiative diffusion efficiently carries the energy flux in deeper regions. The corresponding radius profiles are plotted in panels (c) and (d), with \ce{H2}-rich low MMW models yielding extended envelopes. 
 
Panel (c) shows that assuming a fully-convective structure (dashed lines) acts to deflate the atmosphere, while energy conserving solutions (solid lines) are more inflated for the given $T_s$. In \ce{H2O}-poorest scenarios (red lines), radiative heating inflates the atmosphere from $\sim 4000$ to $\sim \SI{7300}{\kilo\meter}$ ($+83\%$), whereas the \ce{H2O}-richest case is inflated by $+33\%$. Gas escape rates are closely linked to atmospheric radius through equation \ref{eq:EL_escape}, so our radiative-convective model captures the realistic enhancement to photoevaporation here.
 
Panel (b) shows real-gas temperature profiles corresponding to the ideal-gas equivalents in panel (a). Negligible differences between panels show that the EOS formulation only marginally impacts the temperature structure. The choice of EOS matters only when a simplified adiabatic temperature structure is assumed (compare dashed lines between panels c and d) because adiabatic profiles enter into a non-ideal regime. Radiative-convective solutions remain sufficiently hot for the gas to behave ideally. Correspondence between our real and ideal EOS formulations, where expected, demonstrates that our real-gas EOS formulation is behaving correctly.
 
We employ realistic modelling of atmospheric temperature and radius structure, which ensures energy conservation through each model layer. Although Supplementary~Figure~\ref{fig:si_struct} shows that the ideal gas EOS can be safely applied, we choose to employ our real gas EOS formulation for completeness.

\begin{suppfigure}
    \centering
    \includegraphics[width=\linewidth, keepaspectratio]{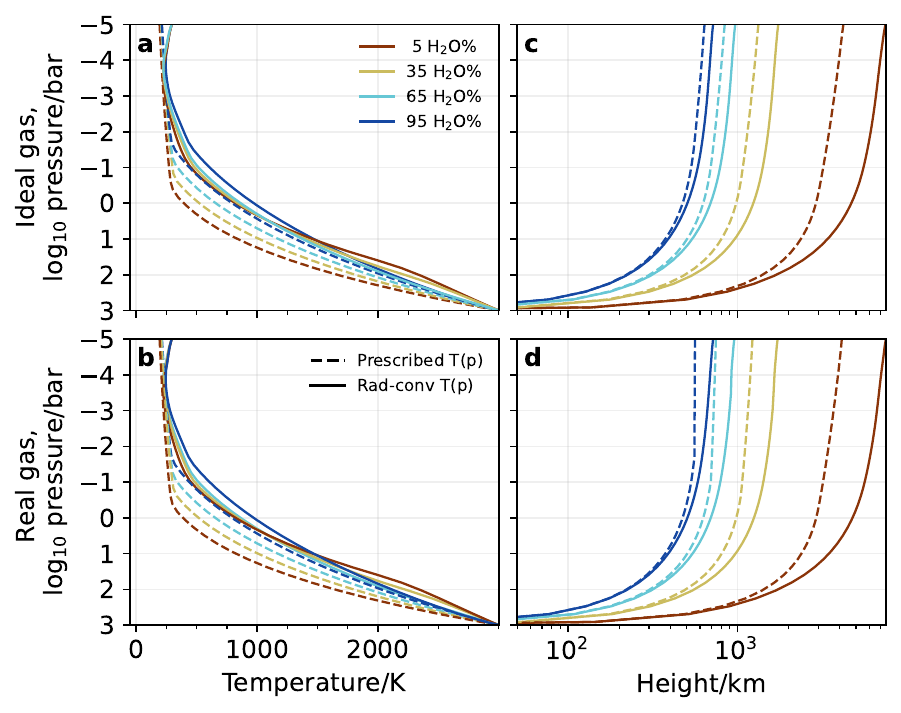}%
    
    \caption{\textbf{Atmospheric temperature (left) and radius (right) structures calculated using AGNI for a range of \ce{H2}-\ce{H2O} mixtures (line colours)}. Panels (a) and (c) use the ideal gas EOS, while panels (b) and (d) use the real gas EOS formulation described in the text. Line style indicates whether a prescribed or energy-conserving temperature solution was obtained.}
    \label{fig:si_struct}
\end{suppfigure}

\subsection{Inclusion of additional volatile species}
\label{sec:si_chem}

Considering the observational indications of \ce{H2S} in the atmosphere of L\,98-59\,d, we update our outgassing scheme \citep{nicholls_redox_2024} to include \ce{H2S}. The partial pressure formed by the exothermic net reaction \ce{S2 + 2 H2 = 2 H2S} is determined by $K_\text{eq}$ at thermochemical equilibrium. Given the relation $K_\text{eq} = \exp(-\Delta G / RT)$, where $\Delta G$ is the reactional change in Gibbs free energy, this equation lets the model account for the speciation of sulfur and hydrogen into \ce{H2S} within the atmosphere overlying the magma ocean. The \textsc{NIST-JANAF} database tabulates $\Delta G$ for \ce{S2} and \ce{H2S} as a function of temperature. The equilibrium constant of this reaction is fitted as $K_\text{eq}(T) = \exp(a/T - b)$ where $a=6731.02 \text{ K}$ and $b=3.62273$, valid from \SI{200}{} to \SI{4000}{\kelvin}.  This reaction is exothermic, so \ce{H2S} is expected to form favourably at surface temperatures near or after mantle solidification. Consistency between lime-coloured \ce{H2S} profiles in Figure~\ref{fig:photochem} shows that this net reaction independently reproduces the same behaviour as \textsc{vulcan}'s full chemical network.

Hydrogen-rich atmospheres could also allow for the formation of \ce{NH3}. Non-detection of \ce{NH3} in the sub-Neptune K2-18 b is central to the ongoing discussion as to its surface conditions, as \ce{NH3} could be readily dissolved in either a magma ocean or water ocean \citep{shorttle_k218b_2024,madhusudhan_carbon-bearing_2023}. We also include the formation of ammonia through the reaction \ce{3 H2 + N2 = 2 NH3}. This reaction is well-fitted with $a=2664.02 \text{ K}$ and $b=5.99238$.

\subsection{Bracketing our parameter space}
\label{sec:si_degas}

The planet's \textit{initial} inventory of volatiles is defined in our model by:
\begin{itemize}
    \item $\Hppmw$, the total mass of hydrogen relative to the mass of the mantle.
    \item S/H, the amount of sulfur in the planet relative to H by mass.
    \item N/H, the amount of nitrogen in the planet relative to H by mass.
    \item C/H, the amount of carbon in the planet relative to H by mass.
\end{itemize}
The first two variables are incorporated into our main grid of simulations. Our grid also considers a range of oxygen fugacities $f\ce{O2}$, which are quantified in $\log_{10}$ units relative to the temperature-dependent iron-w\"ustite buffer. To maintain a computationally feasible parameter space, the C/H and N/H are fixed based on Earth's primitive mantle \citep{wang_elements_2018}. Defining initial SNC abundances relative to H means that $\Hppmw$ acts as a measure of the initial \textit{total} volatile endowment. All other quantities (e.g. mantle melt-fraction, surface temperature) are determined during the evolution by other modelled physics.
 
We can expect that this planet contained relatively more H than Earth's primitive mantle did after boil-off, given multiple observational indications of a present-day \ce{H2}-dominated atmosphere. A wide range of bulk H mass inventories are covered by our main grid: from 1000\,ppmw (approximately ten times that estimated for Earth's primitive mantle; \citep{wang_elements_2018}) up to 16000\,ppmw (i.e. 1.6\,wt\% relative to the mantle mass). The upper limit is a numerically-tractable value informed by other modelling \citep{tang_cpml_2024,lopez_understanding_2014}. Recent modelling of the core-powered mass loss of primary envelopes composed only of H/He has suggested that no more than 1.79\,wt\% of hydrogen (relative to the total mass) may be retained following early boil-off, for a planet with mass $2.4\text{ M}_\oplus$ and instellation $3\times\text{Earth's}$ \citep{tang_cpml_2024}. In our main results, the modelled mantles typically correspond to 68\% of the planet's total mass; this theoretical upper limit of 1.79\,wt\% total hydrogen then corresponds to $\Hppmw \sim 26000$ relative to the mass of the \textit{mantle}. 

Although Earth's upper mantle oxygen fugacity has remained nearly constant for billions of years \citep{trail_hadean_2011}, oxygen fugacity may evolve over time -- in particular during the magma ocean stage -- due to several simultaneous processes that modulate the redox state of iron \citep{mccammon_perovskite_1997, wade_core_2005, itcovitz_reduced_2022, kite_water_2021, lichtenberg_redox_2021, krissansen_erosion_2024, Schaefer2024}. To represent the potential diversity implied by these competing mechanisms, we fix oxygen fugacity $f\ce{O2}$ at the magma ocean surface throughout each simulation and \textit{broadly} bracket a wide range $f\ce{O2}$ by comparing against observations \citep{gressier_hints_2024}. The \SI{9.18}{\amu} contour in Supplementary~Figure~\ref{fig:si_outgas}, equal to the molar mass inferred from retrievals, broadly rules out $f\ce{O2}<\IW-4$ and $f\ce{O2}>\IW$. This suggests that L\,98-59\,d must have an interior much more reducing than Earth's mantle ($\sim\IW+3.5$; \citep{nicklas_redox_2018}) but more oxidising than Mercury's surface ($\sim\IW-5.4$; \citep{namur_mercury_2016}). This stand-alone calculation does not account for changes in abundances induced by escape, which are accounted for in our evolutionary models.
 
S/H is not well constrained by Supplementary~Figure~\ref{fig:si_outgas} at reducing conditions; all S/H are compatible with the retrieved MMW and \ce{H2S} depending on magma ocean oxygen fugacity. \ce{SO2} remains a minor constituent of the atmosphere throughout the phase space as speciation of S into \ce{H2S} is preferred (see photochemical production in Figure~\ref{fig:photochem}). Abundances of nearby stars and models of protoplanetary disk chemistry together indicate that a range of formation scenarios can reasonably allow rocky planets to form with more than 2\% sulfur by mass, with formation scenarios exterior to the ice line generally not exceeding 10 wt\% \citep{jorge_forming_2022, oosterloo_effects_2025}. 

In summary: applying Supplementary~Figure~\ref{fig:si_outgas} alongside reasoning from established cosmochemistry and planet formation constraints provides broad bounds on the initial volatile inventory. We construct our main grid of simulations with four axes: 
\begin{itemize}
    \item oxygen fugacity, $\IW-4.5 \le f\ce{O2} \le \IW$
    \item hydrogen inventory, $1000 \le \Hppmw \le 16000$ 
    \item bulk S/H mass ratio, $2 \le \text{S/H} \le 10$ 
    \item total planet mass, $1.85 \le M_p \le 2.39$ 
\end{itemize}
Mass is varied across the $\pm1\sigma$ range estimated by \citep{rajpaul_doppler_2024}. This parameter configuration results in the 900 scenarios modelled in our main Results (Figure~\ref{fig:density_blame}) with total CHNOS inventories between 0.1\% and 10.9\% of the planet's total mass (median of 2.7\%).

\begin{suppfigure}[!hbp]
    \centering
    \includegraphics[width=\linewidth, keepaspectratio]{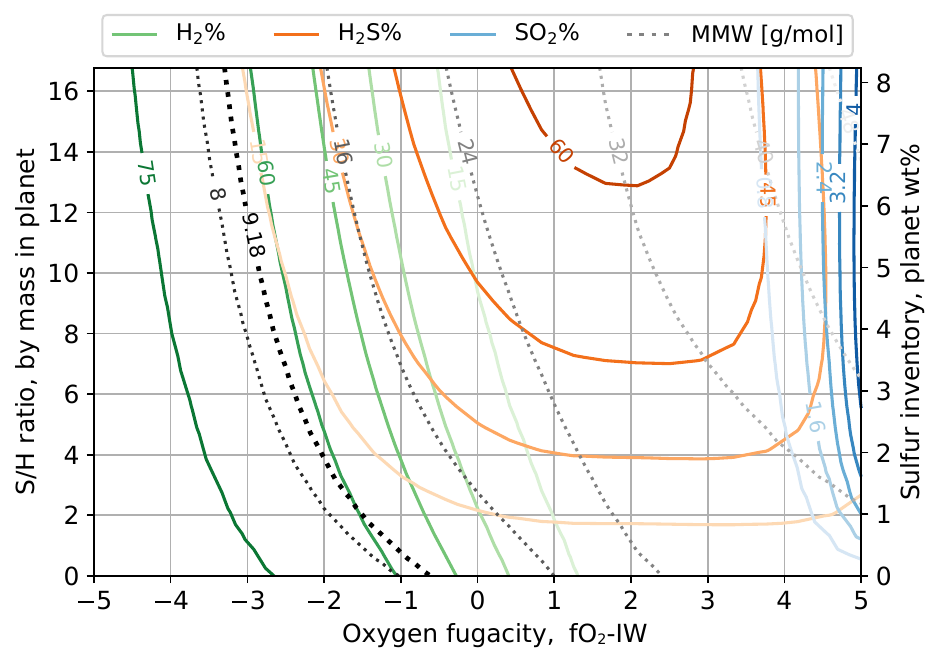}%
    \caption{\textbf{Modelled outgassed atmospheric compositions in equilibrium with an underlying magma ocean}. Coloured contours plot the volume mixing ratios of \ce{H2}, \ce{H2S}, and \ce{SO2} versus bulk S/H mass ratio and oxygen fugacity. Dotted grey lines shows contours of MMW. Outgassing conditions in this plot are \SI{1850}{\kelvin} with a mantle melt-fraction at 40\%, representative of a planet kept semi-molten by tidal heating and an atmospheric greenhouse, $\Hppmw\approx7150$, and $M=2.14 M_\oplus$.}
    \label{fig:si_outgas}
\end{suppfigure}

\subsection{Sensitivity tests to model parameters}
\label{sec:si_sens}

In this section we test the sensitivity of our model to several parameters not included within our main grid. We take one of the configurations from the grid which reproduces an estimate of the present-day bulk-density and atmospheric MMW as a base-case, and then vary: orbital eccentricity $e$, escape efficiency $\eta$, metallic core radius fraction $r_c$, and the number of radiative transfer spectral bands. For these tests, we adopt a baseline configuration with fixed hydrogen content equivalent to $\Hppmw=16000$, core fraction $r_c=55\%$, S/H=8, $f\ce{O2}=\IW-2.5$, and total mass $M_p=2.14\text{ M}_\oplus$.

\begin{suppfigure*}[htbp]
    \centering
    \includegraphics[width=0.9\textwidth, keepaspectratio]{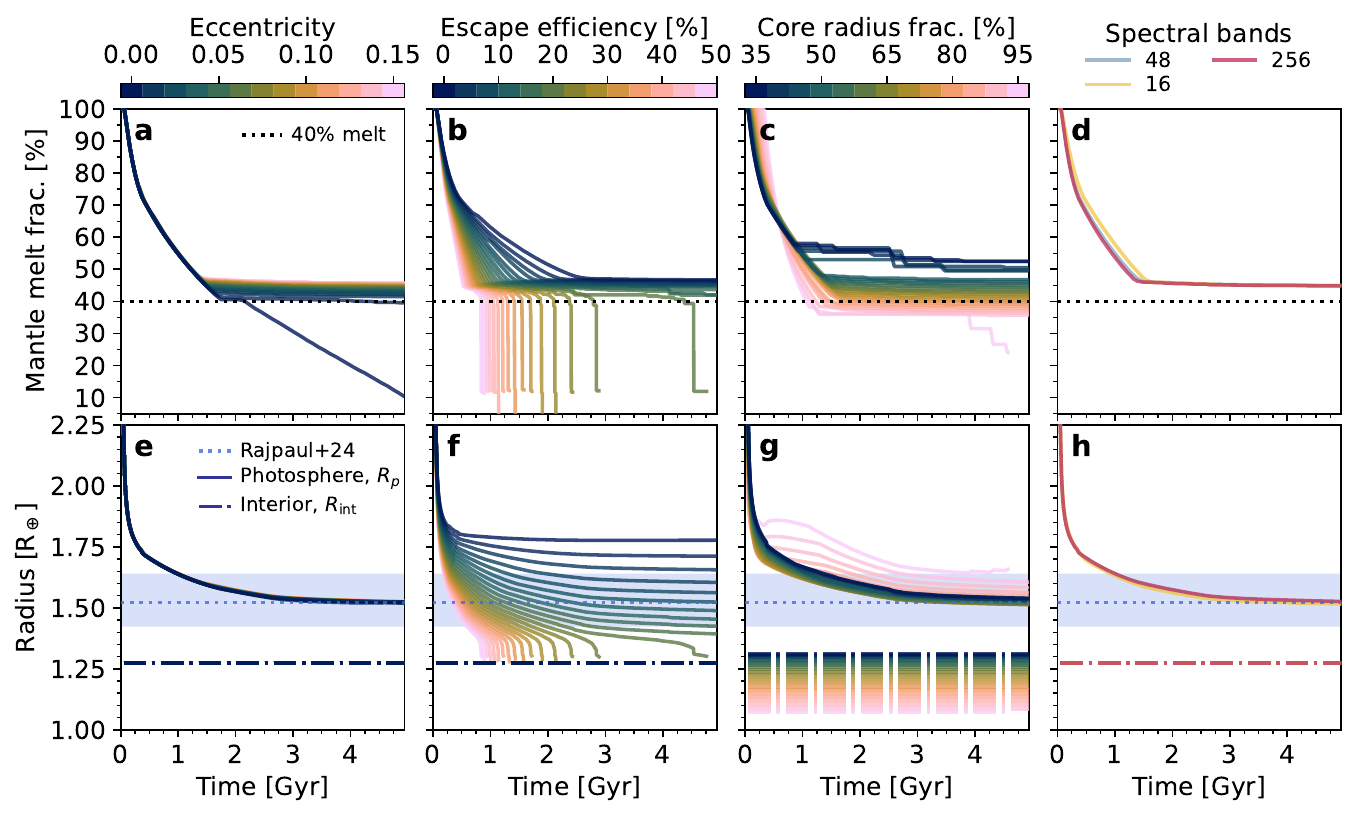}%
    \caption{\textbf{Sensitivity of modelled mantle melt-fraction and effective photospheric radius to model parameters}. Model parameters (columns): orbital eccentricity, atmospheric escape efficiency, metallic core radius fraction, and the number of spectral bands. Sensitivity is assessed by the evolutionary differences in modelled mantle melt-fraction (top row) and the effective photospheric radius $R_p$ of the planet (solid lines in bottom row). Horizontal dashdot lines show interior radius $R_\text{int}$, which varies as a function of core fraction in panel g but is constant in time.}
    \label{fig:si_sens}
\end{suppfigure*}

Sensitivity tests plotted in Supplementary~Figure~\ref{fig:si_sens} show that our main conclusions are robust to variations in these parameters. Panels (a) and (b) reveal a weak dependence on orbital eccentricity; only the exactly-zero eccentricity case tends towards complete mantle solidification. Orbital evolution, not modelled here, would correspond to larger initial eccentricities and an initially increased tidal heating rate \citep{driscoll_tidal_2015}. 

Changes to escape efficiency $\eta$ have the largest impact on modelled evolution (panels b and f) across these four test suites. Increasing $\eta$ to 20\% leads to smaller radii and melt-fractions. Escape efficiencies of $\gtrsim20\%$ -- unrealistic due to the effects of molecular line cooling \citep{yoshida_escape_2024} -- cause the planet to lose its volatile envelope in this case (see next Supplementary Information section).

Panels (c) and (g) show that the modelled thermal outcome of planetary evolution has only minor sensitivity to core radius fractions $r_c$ for $r_c$ up to $80\%$. Since varying core size leads to variation in the mantle mass for a given total planet mass $M_p$, we choose to absolutely specify the initial H inventory as H=914~Earth~oceans for the sensitivity test in panels (c) and (g); this is equivalent to our nominal $\Hppmw=16000$ at a core fraction $r_c=55\%$. However, it should be noted that core size and the H content of the mantle could be physically related through the processes of planet formation and core segregation \citep{peslier_water_2017, wade_core_2005}. Supplementary~Figure~\ref{fig:si_sens}g shows that planet's interior radius $R_\text{int}$ (dashdot lines in bottom row) decreases for increasing $r_c$ since the core material is denser than that of the mantle \citep{bower_numerical_2018}. Cases with extremely large cores, $r_c >80\%$, correspond to shallow mantles hosting magma oceans which are readily saturated in volatiles. So, although H is generally partitioned into the atmosphere under these reducing conditions, small mantles in the case of extremely large cores yield larger effective photospheric radii (solid lines) given the fixed H inventory in this test.

A core compressed to $\rho_c=\SI{15.3}{\GPCC}$ by overburden forces \citep{Hirose2021} would have a radius fraction of $r_c=49\%$ for the same mass as our nominal core configuration of $r_c=55\%$ with $\rho_c=\SI{10.738}{\GPCC}$ based on the PREM \citep{bower_numerical_2018}. Minor differences between $r_c$ of 55\% and $\sim50\%$ in our Supplementary~Figure~\ref{fig:si_sens}c then indicates that holding $\rho_c$ constant in time does not have a substantial significant impact on the thermal evolution modelled in this work.

Panels (d) and (h) show that our nominal choice of 48 correlated-$k$ spectral bands does not bias the calculated observable radii and modelled thermal state of the planet's interior, quantified by its mantle melt fraction.

\subsection{Escape efficiency versus volatile endowment}
\label{sec:si_maxesc}

From the previous section, Supplementary~Figure~\ref{fig:si_sens}f points to some of our modelled outcomes potentially being sensitive to the nominal choice of 10\% escape efficiency $\eta$. The total volatile loss, and thus $\eta$, is influenced by several uncertainties: the star’s rotational evolution and high-energy spectrum \citep{johnstone_active_2021}, and variations in escape rates from three-dimensional star-planet interactions and non-LTE cooling processes \citep{GronoffReview}. However, we posit that sensitivity to $\eta$ can be offset by variations in the planet's initial volatile inventory. To illustrate this, we run an additional grid of simulations which span a plausible range of these parameters: $8000 \le \Hppmw \le 30000$ and $5\% \le \eta \le 25\%$, also varying $f\ce{O2} = \{\IW-2,\IW-3\}$ for completeness. A plausible theoretical value of this planet's H inventory after boil-off is $\sim26000 \ppmw$ \citep{tang_cpml_2024}.

Contours of modelled bulk-density versus $\Hppmw$ and $\eta$ are plotted in Supplementary~Figure~\ref{fig:si_maxesc} for two values of oxygen fugacity (panels a, b). The bulk density of $\rho_p=3.45^{+0.59}_{-1.00}$  estimated by \citep{rajpaul_doppler_2024} is indicated by the blue contours. Diagonal bands show that larger escape efficiencies necessitate larger initial volatile inventories in order to obtain the estimated present-day $\rho_p$. Taking an enhanced 15\% efficiency, following the blue line in panel (a), $f\ce{O2}=\IW-3$, shows that an initial volatile inventory $\Hppmw=26000$ still results in a modelled present-day $\rho_p$ that is consistent with \citet{rajpaul_doppler_2024}. Uncertainties in the observed density (dashed lines) are a more significant factor than our choice of $\eta$\ -- and recent estimates suggest even lower $\rho_p$ for this planet. Considering the $\pm1\sigma$ contours (dashed lines), both large ($\eta\approx20\%$) and small ($\eta\approx8\%$) escape efficiencies yield simulation outcomes within the observational constraints for reasonable volatile-rich scenarios. With a less reducing mantle at $\IW-2$, panel (b), escape efficiencies of 20\% result in present-day $\rho_p$ falling within the observational uncertainty for a reasonable range of $\Hppmw$. These tests show that our main conclusions are robust since all $\Hppmw$ modelled in Supplementary~Figure~\ref{fig:si_maxesc} represent volatile-rich formation scenarios. 

\begin{suppfigure}[htbp]
    \centering
    \includegraphics[width=\linewidth, keepaspectratio]{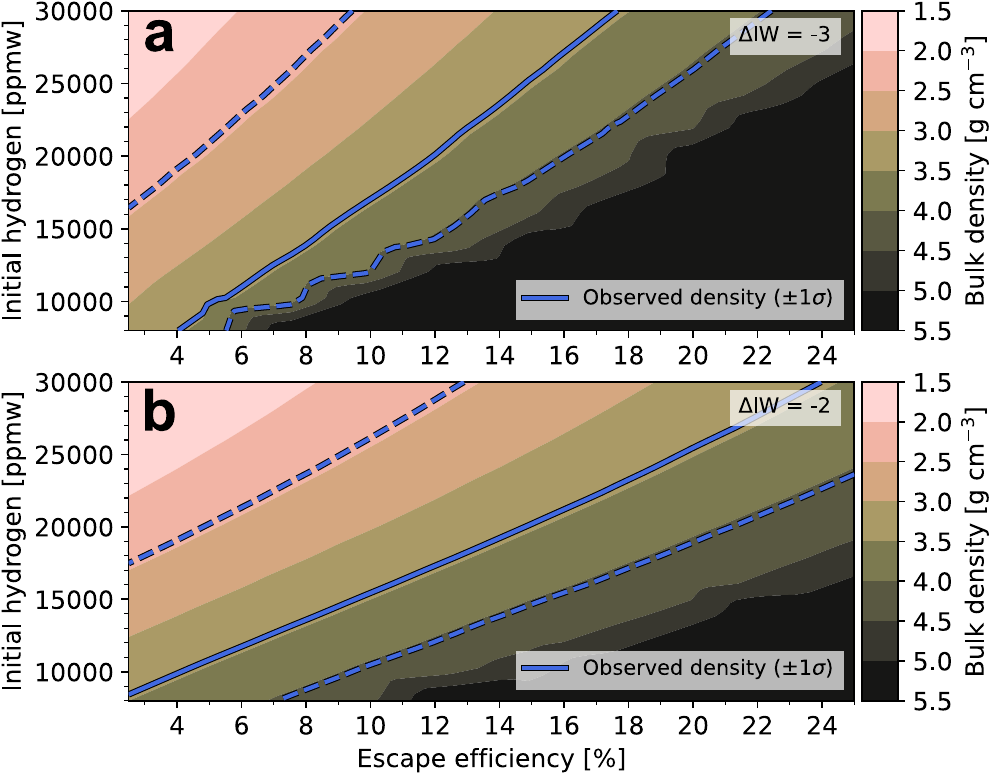}%
    \caption{\textbf{Contour plots of bulk density versus initial volatiles and escape efficiency shows a trade-off between these model parameters}. Modelled planet bulk-density at its present age is indicated by the colour-bar. The density contour estimated by \citet{rajpaul_doppler_2024} is shown by the blue line (solid) with its uncertainty (dashed). Calculated for two values of oxygen fugacity (panels a and b).}
    \label{fig:si_maxesc}
\end{suppfigure}

\subsection{Which physics drives the evolution of bulk planetary characteristics?}
\label{sec:si_driver}

Our simulations show that cooling of a hot interior and atmosphere is the primary driver of initial atmospheric contraction (Figure~\ref{fig:casestudy}b). A decreasing atmospheric height over time corresponds to a decrease in the radius $R_p$ observed by transmission measurements, and corresponds to an increase in $\rho_p$ over time. Figure~\ref{fig:casestudy}b illustrates with pink arrows the two regimes in which atmospheric height decreases over time: first primarily by thermal evolution, and later due to photoevaporative hydrodynamic mass-loss. However, both processes act simultaneously.

To quantify how these different physical processes act to shape planetary evolution, in Supplementary~Figure~\ref{fig:si_driver} we compare variables which trace the physics driving changes in observable properties (solid lines). These correspond to the same simulation as Figure~\ref{fig:casestudy}. Thermal evolution is demonstrated to map to a decrease in atmospheric height, and thus planetary radius (purple and green lines in Supplementary~Figure~\ref{fig:si_driver}). This behaviour is due to the atmospheric scale height decreasing with temperature, through the temperature-dependence of the gas EOS. Furthermore, the partial-solidification of the mantle (brown line) also drives a decrease in $R_p$ for three reasons: (i) solidification causes volatiles to be degassed into the atmosphere -- substantially raising its MMW (blue line) and depressing its scale height, (ii) the thermochemical formation of \ce{H2S} and \ce{CH4} is exothermic -- meaning that these species are favourably generated at cooler temperatures at later times, and (iii) the preferential removal of H atoms from the planet leads to an enhancement in heavier volatiles -- because H is the dominant element in the atmosphere.

Cooling is thus the main reason for substantial radius contraction during the first $\sim 1.4$\,Gyr (green line), since photoevaporative mass-loss is unable to decrease in surface pressure against the effects of volatile outgassing during this time (red line). It is only when the planet attains a thermal steady state (after $\sim1.4$\,Gyr) that the surface pressure decreases over time due to photoevaporative mass-loss, which leads to a slow decrease in planet radius (green line) towards the observational estimate.

Overall, Supplementary~Figure~\ref{fig:si_driver} shows that both thermal evolution (cooling) and compositional evolution (escape and outgassing) have shaped this planet towards the state observed at the present. Different physical processes dominate during each phase of the planet's lifetime, which highlights the importance of comprehensively tracing the Gyr-scale evolutionary behaviour of planets when comparing physical predictions to telescope observations.

Supplementary~Figure~\ref{fig:si_driver} shows a substantial increase in atmospheric MMW (blue line) as volatiles are degassed from the mantle and escape preferentially removes H atoms from the planet. This evolution in MMW is not solely due to photoevaporation, as we model mass-loss as a bulk escape process which is not in-itself fractionating (see \ref{sec:met}). However, since the escaping outflow equals the \textit{atmospheric}  elemental composition, volatile loss fractionates the planet's \textit{bulk} volatile inventories via the different interior-atmosphere partitioning of CHNOS elements. It is therefore important to understand how atmospheric MMW varies more generally -- beyond this case study -- and to understand how changes in atmospheric MMW correlate to changes in mantle melting state, and to changes in the effective radius of the planet over time. Since all of these quantities vary, we compare them by plotting empirical isochrones in Supplementary~Figure~\ref{fig:si_isochrone}, derived from our main grid of models (Section \ref{sec:body}). 

The isochrone at $t=0$ (blue points) corresponds to an initial planet age of 50\,Myr, at which point the mantle is assumed to be entirely molten. The initially low MMW and high temperature yields a large planetary radius (panel a), as expected from the scale height relation. At intermediate times near $t\approx1.4$\,Gyr (orange points), many of these cases approach a thermal steady state where the mantle is kept semi-molten (panel b) by atmospheric blanketing and tidal heating (as per Figure~\ref{fig:casestudy}). The cooling leads to outgassing of heavier elements and the thermochemical formation of heavier species \citep{kite_exchange_2016, nicholls_redox_2024}, which increases the spread of MMW across the models. Cases which solidify (panel b, green points near $\Phi=0\%$) have a large spread of MMWs, up to a maximum of 64\,g/mol where the atmosphere is dominated by \ce{S2} due to the removal of all volatile elements from the planet while S is preferentially retained.

Overall, Supplementary~Figure~\ref{fig:si_isochrone} isochrones show that atmospheric MMW varies substantially in time across a range of possible scenarios. A large radius is generally correlated with a small MMW, but it is important to consider that a range of physics factors into the radius $R_p$ probed by transmission measurements. Maintenance of a semi-molten state restricts the spread of atmospheric MMW, since it means that atmospheric composition is buffered by dissolution of volatiles into the underlying melt.

\begin{suppfigure}[!hbp]
    \centering
    \includegraphics[width=\linewidth, keepaspectratio]{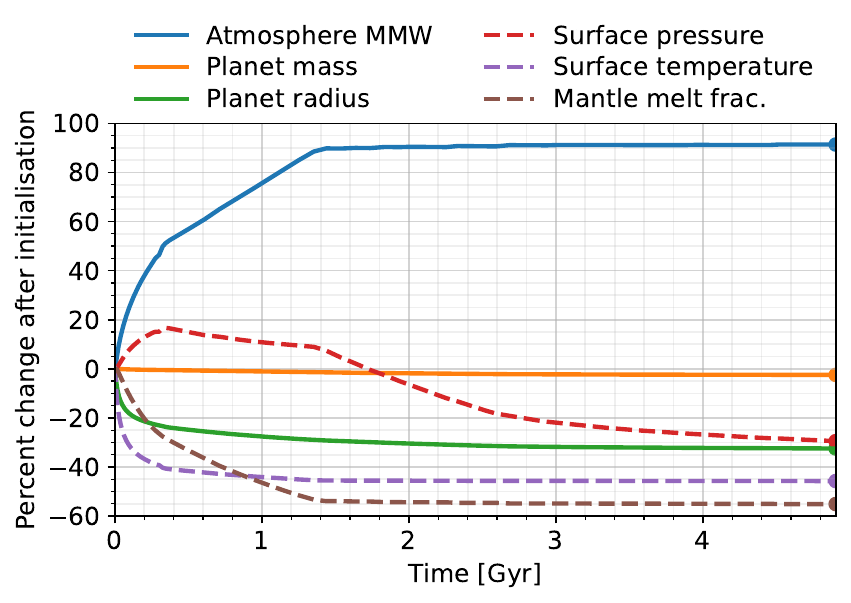}%
    \caption{\textbf{Evolution of modelled variables shows that various physical processes drive an overall decrease in planetary radius}. Coloured lines trace the relative change of six variables, starting from the point of model initialisation at 50\,Myr after planet formation, up to the estimated 4.94\,Gyr present-day age of L\,98-59\,d.}
    \label{fig:si_driver}
\end{suppfigure}
\begin{suppfigure}[!hbp]
    \centering
    \includegraphics[width=\linewidth, keepaspectratio]{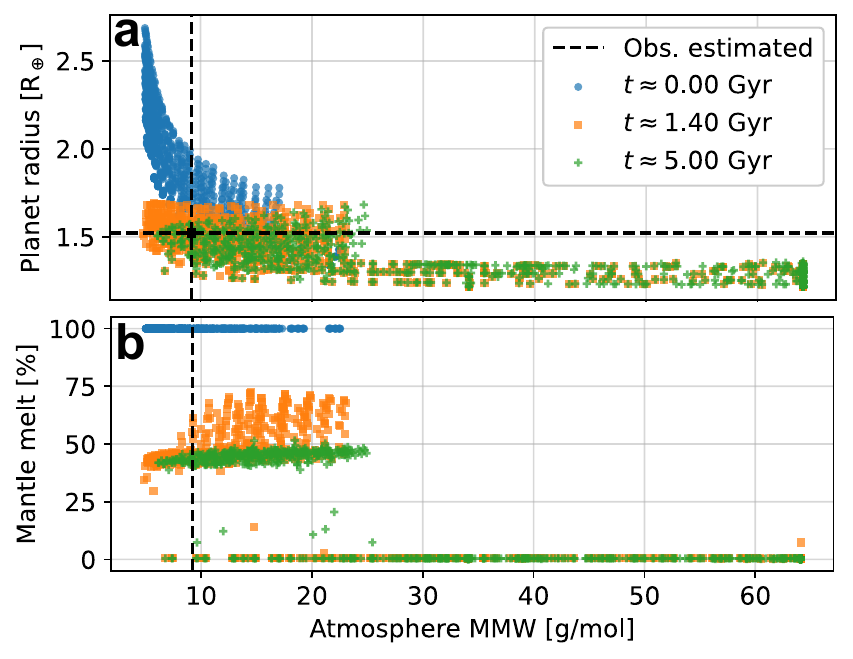}%
    \caption{\textbf{Isochrones of planet radius, melt fraction, and atmospheric mean molecular weight}. Coloured points show the values of these quantities at three points in time, measured relative to model initialisation 50\,Myr after planet formation. }
    \label{fig:si_isochrone}
\end{suppfigure}

\subsection{Robustness to new estimates of planetary and system parameters}
\label{sec:si_update}
A re-analysis of the existing ESPRESSO and HARPS radial velocity measurements of this system -- processed alongside TESS photometry and unpublished JWST transmission spectroscopy -- has recently yielded new estimates for the parameters for L\,98-59 and its transiting planets \citep{cadieux_detailed_2025}. The relevant parameters are listed in Table~1,  including a slower stellar rotation period, and smaller estimates on the mass, radius, and orbital eccentricity of L\,98-59\,d. These differences arise from a different treatment of stellar activity by \citet{cadieux_detailed_2025} when compared to previous works, as well as their joint fit to transit-timing variations alongside radial velocity data. Importantly, the corresponding bulk density of L\,98-59\,d with these estimates is \SI{2.2}{\GPCC}; L\,98-59\,d now represents the puffiest (least dense) known exoplanet in its radius class, highlighting its relevance as a case study for the physics shaping the properties of super-Earths, and the conclusions presented here.

To test the robustness of our main results and conclusions to these new estimates of planetary and stellar parameters, we present an additional set of evolutionary calculations made using \textsc{proteus} in Supplementary~Figure~\ref{fig:si_density_updated} -- considering greater initial volatile inventories and a range of metallic core fractions. While we fix the metallic core radius fraction at 55\% in our main simulations and also perform sensitivity tests to its control over thermal evolution, these new estimates for the bulk properties of L\,98-59\,d fortuitously place L\,98-59\,d within a density regime where its core fraction $r_c$ may be readily constrained by our model. 

\begin{suppfigure}[!hbp]
    \centering
    \includegraphics[width=\linewidth, keepaspectratio]{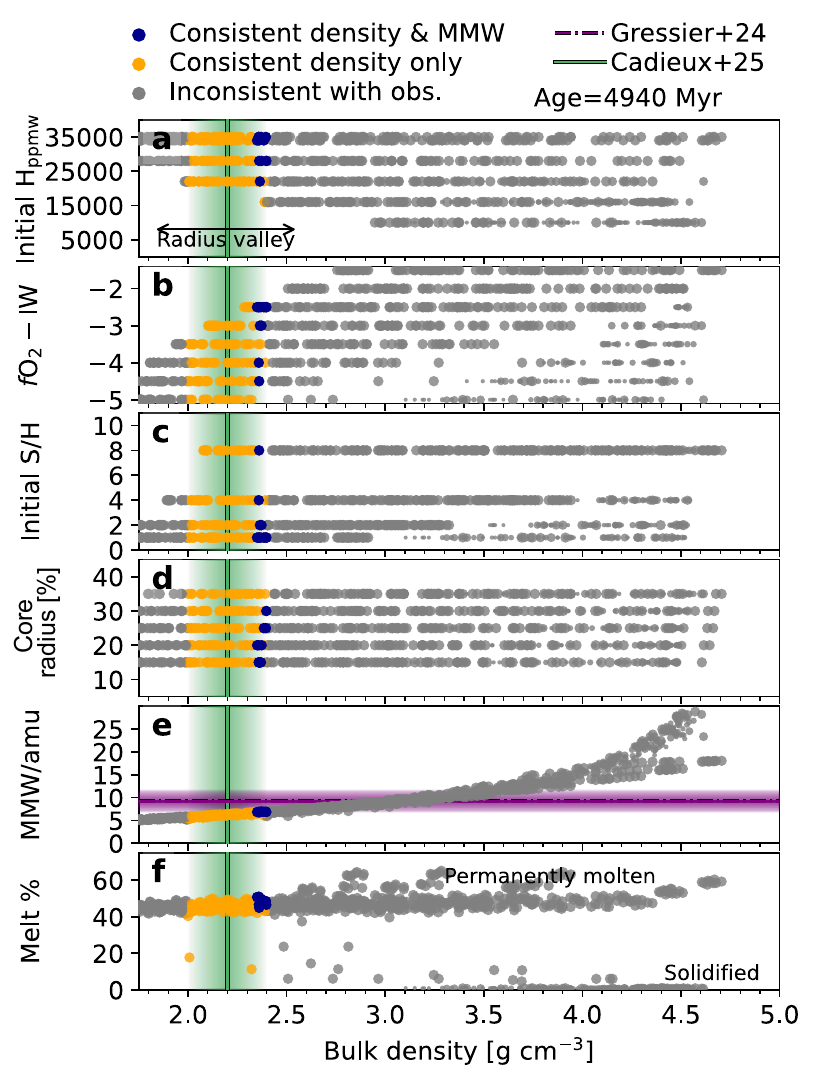}%
    \caption{\textbf{Projection of modelled bulk-density $\rho_p$ against several variables}. Vertical green line and shaded region ($\pm1\sigma$) highlights the recent estimates on $\rho_p$ by \citet{cadieux_detailed_2025}. The horizontal purple line shows estimates on MMW (with $\pm1\sigma$ shaded). New models consistent with the new estimates of the density and the MMW are shown by blue points; cases consistent only with the new estimate of density are shown by orange points.}
    \label{fig:si_density_updated}
\end{suppfigure}

Supplementary~Figure~\ref{fig:si_density_updated} presents a comparison between these our simulations (scatter points) and the recent estimates of planetary bulk density (green region; \citep{cadieux_detailed_2025}). In order to theoretically reproduce this estimate of the planet's bulk density, regardless of its atmospheric MMW, models require that L\,98-59\,d formed volatile-rich and with a hydrogen inventory $\Hppmw\ge22000$ relative to the mass of the planet's mantle (orange points). This H inventory corresponds to large CNOS inventories, as these elements are defined relative to H at model initialisation (see Methods). Also from the new density constraint alone, the orange points in Supplementary~Figure~\ref{fig:si_density_updated} show that the planet's upper mantle must have an oxygen fugacity $f\text{O}_2 \le \IW-2.5$, consistent with the main results presented in Figure~\ref{fig:density_blame}. These new simulations support our main analysis (Figure~\ref{fig:density_blame}) and reinforce our assessment that this planet formed highly volatile-enriched but with a geochemically-reducing composition. 

Furthermore, a comparison between the additional simulations presented in Supplementary~Figure~\ref{fig:si_density_updated}, and the new estimates of density \textit{alongside} the previously-retrieved atmospheric MMW, constrains the metallic core of L\,98-59\,d to comprise $\le30\%$ of the rocky interior by radius (blue scatter points). While requiring a reasonable envelope mass fraction, a small core is necessary for explaining the low estimate of bulk density by \citet{cadieux_detailed_2025}. A small metallic core is additionally consistent with prior static structure models of this planet presented in the literature \citep[e.g.][]{demangeon_warm_2021, cloutier_character_2019}. Observations indicate that the star L\,98-59 has a low metallicity \citep{vallenari_gaiaDR3_2023, demangeon_warm_2021} which could be commensurate with a subsequently small metallic core fraction within the planet L\,98-59\,d. 

The estimate of a 77.6\,day stellar rotation period by \citep{cadieux_detailed_2025} is comparable with the 80.9\,day period modelled in our main results section. This new estimate comes from a different modelling treatment of stellar photospheric temperature in their retrieval framework compared to previous estimates \citep{rajpaul_doppler_2024, demangeon_warm_2021}. In adopting this slower rotation period within these the evolutionary models presented in Supplementary~Figure~\ref{fig:si_density_updated}, we demonstrate that our conclusions are robust to uncertainties in the star's modelled rotational evolution. It should be noted that the rotation rate factors into our modelling only through its impact on the XUV flux, and thus on the escape rate (Equation \ref{eq:EL_escape}). The effect of variations in stellar rotation rate are therefore degenerate with the escape efficiency, for which we perform sensitivity tests in Supplementary~Figure~\ref{fig:si_sens} and \ref{fig:si_maxesc}.

\section*{Data availability}
The simulation data and simulation codes underlying this article are available on Zenodo: \cite{nicholls_zenodo_2025}.

\section*{Code availability}

In addition to the archive on Zenodo, all computer codes used in this work are open source software available on GitHub: 
\begin{itemize}
    \item \textsc{proteus}, \url{https://github.com/FormingWorlds/PROTEUS} 
    \item \textsc{agni}, \url{https://github.com/nichollsh/AGNI} 
    \item \textsc{spider}, \url{https://github.com/djbower/spider}
    \item \textsc{lovepy}, \url{https://github.com/nichollsh/lovepy} 
    \item \textsc{socrates}, \url{https://github.com/nichollsh/SOCRATES}
\end{itemize}

\section*{Acknowledgements}
TL acknowledges support from the Netherlands eScience Center (PROTEUS project, NLESC.OEC.2023.017), the Branco Weiss Foundation, the Alfred P. Sloan Foundation (AEThER project, G-2025-25284), and NASA’s Nexus for Exoplanet System Science research coordination network (Alien Earths project, 80NSSC21K0593). CMG is supported by the UK STFC (ST/W000903/1). RTP and RDC acknowledge support from the UK STFC and the AEThER project. RDC also thanks UKRI for their support via grant number UKRI1191. HN acknowledges support from STFC grant UKRI1184. We thank the Center for Information Technology of the University of Groningen for their support and for providing access to the Habrok high performance computing cluster. We also thank Tad Komacek for his suggestions on this manuscript, and Laurent Soucasse for his contributions to \textsc{proteus}.

\section*{Author contributions}

\textbf{HN:} Conceptualization, Methodology, Software, Investigation, Writing - Original Draft, Writing - Review \& Editing, Visualization.
\textbf{TL:} Conceptualization, Methodology, Software, Writing - Review \& Editing.
\textbf{RDC:} Conceptualization, Methodology, Writing - Original Draft.
\textbf{CMG:} Conceptualization, Writing - Review \& Editing.
\textbf{EP:} Software, Writing - Review \& Editing.
\textbf{RTP:} Supervision.

\pagebreak
\newpage
\bibliography{refs}{}
\bibliographystyle{aasjournal}

\end{document}